# Secrecy Capacity Analysis of 4-WFRFT Based Physical Layer Security in MIMO System

Tao Xu. Author, Xuan li WU. Author

*Abstract*—The traditional information security based on cryptosystem is seriously threatened due to the exponential growth of computing capacity. In order to improve for the upper cryptosystem security, the secure transmission at the physical layer is introduced into the wireless communication system. However, considering the openness of wireless channel, the performance analysis of security wireless communication systems has become a hot issue in recent years. Due to the existence of the upper layer cryptosystem being cracked and the openness of wireless channel security problems, the Multiple-Input Multiple-Output (MIMO) technology increases the difference between channels, the development of its rich spatial degrees of freedom can greatly improve the channel capacity and achieve the purpose of enhancing the Physical layer security (PLS); On this basis, Weighted Type Fractional Fourier Transform (WFrFT) technology rotates and splits the constellation points of signals, it is equivalent to increasing the artificial noise to enhance the PLS. The Average security capacity is an important index to measure the PLS, Therefore, this paper deduces the Average security capacity in MIMO scenario and gives a specific closed expression when considering the channel correlation. By analyzing the expression, the increase of the number of MIMO antennas will improve the Average security capacity. When the estimated bias $\Delta\alpha \neq 0$ in 4-WFRFT, the Average security capacity will be significantly improved; when $\Delta\alpha \in (1,3)$, the Average security capacity will reach the maximum value, thus enhancing the PLS.

## I. INTRODUCTION

The rapid development and wide application of wireless mobile communication technology have brought great convenience to human life. However, due to the openness and broadcasting nature of wireless communication system, any receiver within the communication range can intercept the communication signals, which seriously threatens the secure transmission of information. Therefore, information security becomes an important issue that scholars in this field focused on [1]-[3].

The traditional security mechanism in the field of wireless communication system is an encryption framework based on cryptography theory. The encryption system ensures the security of the communication system by implementing different authentication and encryption algorithms. The concept of this system inherits the traditional computer network and assumes that the physical layer can provide an error-free transmission link, however, it neglects the characteristics of the physical layer in wireless communication system, such as the openness of channel and the time-variability of network topology [4]. However, with the rapid development of computer technology, the computing power of computers has been greatly improved, and the security guarantee provided by the complex algorithm based on the traditional encryption system is increasingly insufficient for the wireless communication system. How to use physical layer characteristics to provide effective security for wireless communication has become a new research hotspot [5,6].

Physical Layer Security (PLS) is absolute security based on information theory. The earliest security theory of information theory can be traced back to 1949 when Shannon put forward the theory of security system [7], i.e., no matter how many resources an eavesdropping has, the secure communication system cannot be cracked. On this basis, Wyner considered the effects of multipath fading and noise in the realistic system and established a three-point eavesdropping channel model [8]. Ref. [9] proposed an optimal beamforming scheme assisted by artificial noise to improve the PLS. In ref. [10], a weighted fractional Fourier transform (WFrFT) precoding scheme is proposed to enhance the security of wireless transmission using the physical layer characteristics. Considering the PLS model of MIMO two-way relay cooperative communication network, ref. [11] designed an allocation optimization algorithm to improve the total security capacity of the system.

Multiple Input Multiple Output (MIMO) technology has attracted wide attention due to its advantages in improving spectrum efficiency, increasing the difference between channels, utilizing rich space resources, and improving the channel capacity significantly. By using the spatial freedom of MIMO into PLS, the security performance can be further improved. The capacity of MIMO system has been greatly improved due to the increase in the number of antennas. However, the increase in the number of antennas will inevitably lead to a significant increase in the complexity of the system algorithm, and the hardware cost of each device will also increase greatly, making the system construction and maintenance difficult. Based on the above factors, scholars have proposed Antenna Selection (AS) technology under multi-antenna to improve system capacity and reduce complexity. Ref. [12] and [13] introduced MIMO into the wireless communication system, and analyzed the reasons for MIMO to improve channel capacity after using AS technology, however, the impact of channel correlation was not considered. After considering the channel correlation, ref. [14]-[16] analyzed the system security performance in MIMO scenarios, but no precise expression of the Average security capacity was given.

In recent years, weighted fractional Fourier Transform (WFRFT), a new signal processing method in the transform domain developed based on Fourier Transform (FT) [17], has been proposed, and it has many advantages, such as uniform distribution of the processed signals in the time-frequency domain, anti-jamming and anti-fading properties, and simple engineering implementation. When the WFRFT is applied into PLS, by changing the order of the transform, the effective information obtained by the eavesdropper is almost zero, thus greatly improving the system capacity. Ref. [18]-[20] have applied WFRFT technology into PLS, and analyzed the impact of WFRFT on system security performance from the perspective of bit error rate, however, channel correlation is not considered, and the impact of WFRFT on Average security capacity was not analyzed. In [21,22], the 4-Weighted-Type Fractional Fourier Transform (4-WFRFT) analysis of Average security capacity is given in MIMO scenarios, but the exact expression of Average security capacity is not given.

The existing work analyzed the system security performance in MIMO scenarios and WFRFT, respectively, they gave the factors that affect the system security performance, but didn't derived the specific expression of Average security capacity , however, the specific expression of average security capacity is crucial for analyzing system security performance, giving the average security capacity has great significance. Combined with the existing work, under the premise of considering the channel correlation, the paper comprehensively considers 4-WFRFT technology and the AS technology in MIMO scenario, and derives the closed analytic expression of Average security capacity, and moreover, the parameter selection to improve the Average security capacity is also given.

The contributions of this paper are as follows:

(1) The AS technology and 4-WFRFT technology are combined into the MIMO system, and the influence of these two technologies on the joint PDF of the system channel is derived. The closed analytical expression of Average security capacity under the condition of channel correlation is derived, and its accuracy is verified by simulation.

(2) The influence of system parameters on Average security capacity is obtained through simulation, and the selection basis and corresponding method of relevant parameters (correlation coefficient, number of antennas, transmitting power and transformation parameter) are obtained under the scenario of AS and 4-WFRFT.

The following chapters of this paper are arranged as follows: Section II considers channel correlation in MIMO system, analyzes channel correlation coefficient, and establishes Rayleigh fading channel model; In Section III, in order to enhance the PLS, AS and 4-WFRFT technology are introduced into MIMO system, and the joint PDF of SNR at receiver is given. In Section IV, combined with the PDF, the closed analytic expression of Average security capacity with AS and 4-WFRFT is given. In Section V, the method of improving Average security capacity is obtained by simulation analysis combined with theory. The Section VI gives the conclusion.

## II. SYSTEM MODEL AND PROBLEM FORMULATION

In this Section, we will firstly establish Rayleigh fading channel model and introduce channel correlation coefficient into Rayleigh fading system. Then, system model will be given with two key technologies, AS and 4-WFRFT, and the expression of receiver's SNR problem will be fomulated as well.

### A. Correlated Rayleigh fading channel model

In the realistic communication environment, the mobile receiver is in constant motion, the receiver receives multipath signals from different paths, which is in line with the wireless communication environment with obstacles in the general city or mountain area, without the feature of direct path. Therefore, the system is modeled as Rayleigh fading channel model.

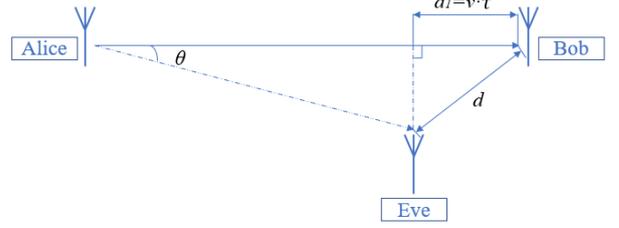

Fig.1. Rayleigh fading eavesdropping channel model

Fig.1 is the modeling of correlated Rayleigh fading channel. In multipath fading channels, channel correlation is related to channel characteristic parameters such as time difference $\tau$, Doppler frequency shift $f_d$ and antenna distance $d$. Alice encodes the signal and sends the generated code word to the legal receiver Bob, while Eve, the illegal eavesdropper, eavesdrops on the information sent by Alice to Bob. Eve is in a passive eavesdropping state and does not interfere with the main channel. The correlation of channel is reflected in the correlation between the channel fading coefficients $h_M$ and $h_E$. In Rayleigh fading channel model, there are multiplicative fading coefficient $h$ and additive Gaussian white noise $n$. Therefore, the signal received by the legal receiver and the eavesdropper is:

$$y = \sqrt{P}h_M x + n_M, \quad (1)$$

$$z = \sqrt{P}h_E x + n_E, \quad (2)$$

where $P$ is the transmitted power of Alice; The channel fading coefficients of $h_M$ and $h_E$ obey the complex Gaussian distribution, which are the main channel and eavesdropping channel, respectively. That are $h_M \sim \mathcal{CN}(0,\sigma_M^2), h_E \sim \mathcal{CN}(0,\sigma_E^2)$; $n_M$ and $n_E$ are the noise of the main channel and the eavesdropping channel respectively, which obey the Gaussian distribution, that is $n_M \sim \mathcal{N}(0,N_M), n_E \sim \mathcal{N}(0,N_E)$. When channel correlation is considered, the channel fading coefficients $h_M$ and $h_E$ can be expressed as:

$$\frac{h_M}{\sigma_M} = \left(\sqrt{1-\eta^2}X_M + \eta X_0\right) + j\left(\sqrt{1-\eta^2}Y_M + \eta Y_0\right), \quad (3)$$

$$\frac{h_E}{\sigma_E} = \left(\sqrt{1-\lambda^2}X_E + \lambda X_0\right) + j\left(\sqrt{1-\lambda^2}Y_E + \lambda Y_0\right), \quad (4)$$

where $X_M$, $X_E$ $Y_M$, $Y_E$, $X_0$ and $Y_0$ are independent of each other and obey the Gaussian distribution of $\mathcal{N}(0,1/2)$; $\sigma_M$, $\sigma_E$ are used to adjust the size of the channel fading coefficient, both $\eta$ and $\lambda$ satisfy $|\eta|<1, |\lambda|<1$ weight coefficients.

Defined by correlation coefficient, the correlation coefficient between $h_M$ and $h_E$ of the channel fading coefficient can be expressed as:

$$\rho = \frac{\text{Cov}[h_M, h_E]}{\sqrt{D[h_M]}\sqrt{D[h_E]}} = \eta\lambda\left\{E\left[X_0^2\right] + E\left[Y_0^2\right]\right\} = \eta\lambda, \quad (5)$$

when $\eta$ and $\lambda$ change, in turn, the channel correlation $\rho$ changes, correlation coefficient describes the main channel and the eavesdropping channel similar degree between the channel.

The receiver instantaneous SNR is expressed as:

$$\gamma_M = P\frac{|h_M|^2}{N_M} = P\frac{\sigma_M^2}{N_M}\left[\left(\sqrt{1-\eta^2}X_M + \eta X_0\right)^2 + \left(\sqrt{1-\eta^2}Y_M + \eta Y_0\right)^2\right], \quad (6)$$

$$\gamma_E = P\frac{|h_E|^2}{N_E} = P\frac{\sigma_E^2}{N_E}\left[\left(\sqrt{1-\lambda^2}X_E + \lambda X_0\right)^2 + \left(\sqrt{1-\lambda^2}Y_E + \lambda Y_0\right)^2\right], \quad (7)$$

the average SNR $\overline{\gamma_E}$, $\overline{\gamma_M}$ as follows: $\overline{\gamma_E} = P\frac{\sigma_E^2}{N_E}$ and $\overline{\gamma_M} = P\frac{\sigma_M^2}{N_M}$.

In the actual communication system, the eavesdropping channel and the main channel can't be regarded as completely independent, and the correlation between the two channels is universal. The degree of correlation largely depends on the communication environment. When the eavesdropper is close to the legitimate receiver for active correlation and the scatterers around it have high similarity, there will be a high degree of channel correlation for receivers.

Channel correlation is divided into temporal correlation and spatial correlation, which are equivalent when the eavesdropper is close to the legitimate receiver. The channel correlation coefficient can be expressed as:

$$\rho = J_0(2\pi f_d \tau) = J_0\left(2\pi \frac{d}{\lambda}\right), \quad (8)$$

where $\tau$ is time difference, $f_d$ is doppler frequency shift, both embodied in time; $d$ is antenna distance, $\lambda$ is signal wavelength, which embodied in space; $J_0(x)$ is the first kind of zero-order Bessel function, defined as: $J_0(x) \triangleq \frac{1}{2\pi}\int_0^{2\pi} e^{-j\cdot x\cos\theta} d\theta$.

### B. System Model

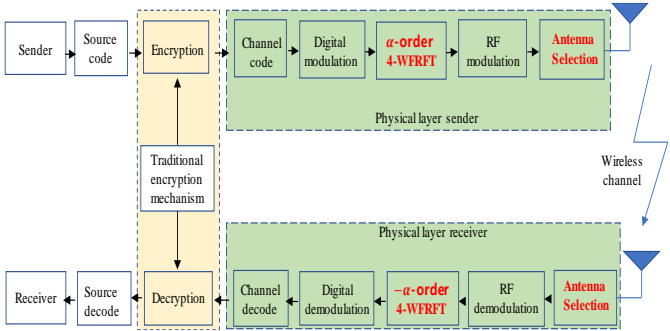

Fig.2 Physical layer security mechanism

Fig. 2 shows the physical layer security mechanism. In order to enhance the reliability of wireless channel transmission, we have added 4-WFRFT and AS technologies as a supplement to the system security in the physical layer security mechanism,

AS technology is introduced in MIMO system, AS shows in Fig.3, the main channel for $N_A \times N_B$ MIMO channels, transmitter Alice configures $N_A$ antenna, legal receiver Bob configures $N_B$ antenna.

Compared with Single input single output (SISO) system, the capacity of MIMO has been greatly improved with the increase of the number of antennas. However, the increase of the number of antennas will bring a significant increase in the complexity of the system algorithm, and the difficulty of system construction and maintenance will be increased. Therefore, researchers hope to propose a reasonable transmission scheme under multiple antennas to improve system capacity and reduce system complexity. Therefore, AS comes into being.

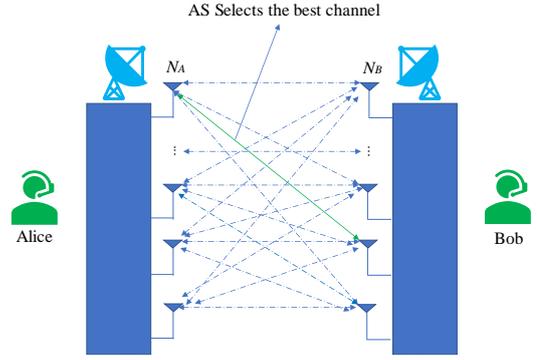

Fig.3. AS applied to MIMO model

AS principle, by Alice with $N_A$ antenna to send pilot signal, Bob with $N_B$ antenna for receiving, through to the antenna position deployment, can make there is no correlation between antenna, produce $N_A \times N_B$ independent fading channels. After receiving the pilot signal, the receiver determines the channel state according to the characteristic parameters such as signal strength, and selects the antenna with the best channel state, and sends the feedback information to the transmitter through the feedback channel and feedback link. The system determines the transmitting and receiving antenna of the channel as communication antennas. However, the eavesdropper doesn't have such a cooperative mechanism, so the AS is invalid for the eavesdropper, so its receiving SNR will not change.

In order to improve the PLS of system, WFRFT technology is introduced. This section considers the influence of 4-WFRFT in WFRFT on the security performance of MIMO. Discrete Fourier Transformation (DFT) was used to compute 4-WFRFT of discrete sequences. For discrete sequence of length $N$: $x(n)(n=0,1,\cdots,N-1) \in C^N$, DFT can be expressed as:

$$X(k) = \frac{1}{\sqrt{N}}\sum_{n=0}^{N-1} x(n)e^{-i\frac{2\pi}{N}nk}, \quad (9)$$

$$x(n) = \frac{1}{\sqrt{N}}\sum_{k=0}^{N-1} X(k)e^{i\frac{2\pi}{N}nk}, \quad (10)$$

4-WFRFT is defined as:

$$\mathcal{L}_{4w}^\alpha[x(n)] = \omega_0(\alpha)X_0(n) + \omega_1(\alpha)X_1(n) + \omega_2(\alpha)X_2(n) + \omega_3(\alpha)X_3(n), \quad (11)$$

where $\alpha$ is transform order number, $\omega_p(\alpha) = \cos[(\alpha-p)\pi/4]*\cos[2(\alpha-p)\pi/4]*\cos[\pm 3(\alpha-p)i\pi/4], (p=0,1,2,3)$; $X_0(n), X_1(n), X_2(n), X_3(n)$ are the 0, 1, 2 and 3 DFT of $x(n)$ respectively. According to DFT, there are the following transformations:

It can be obtained that the result of 4-WFRFT of signal $X_0$ can be expressed as the weighted superposition of $X_0(n)$

$$\mathcal{L}^1[x(n)] = \mathcal{L}^1[X_0(n)] \xrightarrow{n=k} X_1(n) = X(n), \quad (12)$$

$$\mathcal{L}^2[X_0(n)] = \mathcal{L}^1[X_1(n)] \xrightarrow{n=k} X_2(n) = x(-n), \quad (13)$$

$$\mathcal{L}^3[X_0(n)] = \mathcal{L}^1[X_2(n)] \xrightarrow{n=k} X_3(n) = X(-n), \quad (14)$$

$$\mathcal{L}^4[X_0(n)] = \mathcal{L}^1[X_3(n)] \xrightarrow{n=k} X_4(n) = x(n), \quad (15)$$

time domain representation and frequency domain representation. (13) and (14) are inverted functions of time domain signal $x(n)$ and frequency domain signal $X(n)$ centered on the origin respectively.

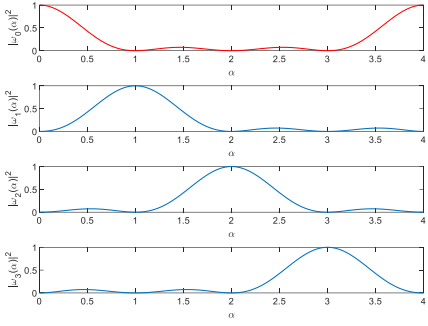

Fig.4. Relationship between $|\omega_p(\alpha)|^2$ and $\alpha$

Fig.4 shows that $|\omega_P(\alpha)|^2$ cycle is 4, and with the changes of $\alpha \in [0, 4)$ within the scope of change, the weight of each component is cyclical change. When $\alpha$ approaches to 0 or 2, $|\omega_0(\alpha)|^2$ and $|\omega_2(\alpha)|^2$ is 1, the time domain component proportion increase; When $\alpha$ approaches to 1 or 3, $|\omega_1(\alpha)|^2$ and $|\omega_3(\alpha)|^2$ is 1, the frequency domain component proportion increased. Especially, when $\alpha = 0$, corresponding to the function of the weighted coefficient $\omega_0(\alpha) = 1$, other coefficient is 0, 4-WFRFT degenerates into the original function; when the $\alpha = 1$, the corresponding Fourier transform component of the weighted coefficient of $\omega_1(\alpha) = 1$, other factors are 0, 4-WFRFT degenerates into the classical Fourier transform. 4-WFRFT be applied to Rayleigh fading channel system, and its related channel model is as follows Fig.6.

Alice and Bob share 4-WFRFT transform order $\alpha$. Transmitter Alice transforms order $\alpha$ for 4-WFRFT, legal receiver Bob transforms order number $-\alpha$ for 4-WFRFT. Eve is eavesdropper, through technical means don't obtain transform order $\alpha$ exact value, but can choose a change order number $-\beta$ as $-\alpha$ estimates, then transforming order $-\beta$ for 4-WFRFT. Here defining estimate bias $\Delta\alpha \triangleq |\alpha - \beta|$.

Assume that $x_k$ is input data after baseband modulation, after 4-WFRFT with the transformation order $\alpha$, the signal becomes $S_k$ and is sent out, satisfy the following relations:

$$S_k = \mathcal{L}^\alpha[x_k], \quad (16)$$

according to the relevant Rayleigh fading channel model established, at this time, the signals received by Bob and Eve can be expressed as:

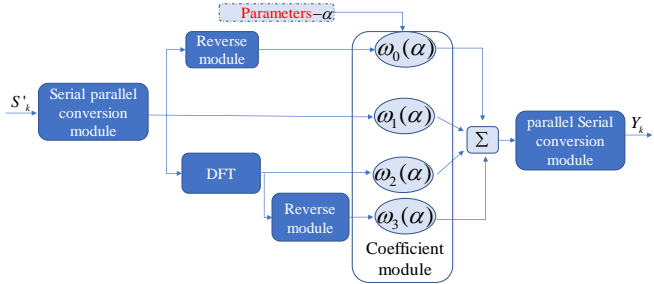

Fig.5. 4-WFRFTA Flow chart

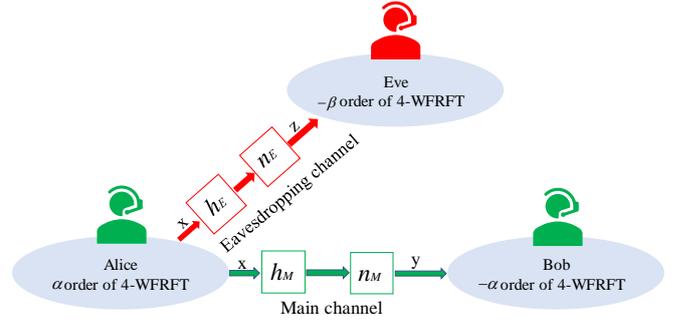

Fig.6. 4-WFRFT applied to Rayleigh fading channel model

$$y_k = \sqrt{P} h_M S_k + n_M, \quad (17)$$

$$z_k = \sqrt{P} h_E S_k + n_E, \quad (18)$$

Since Bob shares transformation order $\alpha$ with Alice, the legitimate receiver can carry out 4-WFRFT with transformation order $-\alpha$, then the signal received by Bob can be further processed as follows:

$$\begin{aligned} r_{Bk} &= \mathcal{L}^{-\alpha}[\sqrt{P} h_M \mathcal{L}^\alpha[x_k] + n_M)] \\ &= \sqrt{P} h_M x_k + n_M \end{aligned}, \quad (19)$$

for Bob, because Alice and Bob share transform order $\alpha$, suggests that Bob's instantaneous SNR $\gamma_M$ is not affected by 4-WFRFT, same as (6).

Eve doesn't know the transformation order $\alpha$ and chooses transformation order $-\beta$ as the estimator of $-\alpha$, to perform 4-WFRFT of transformation order $-\beta$. Then the signals received by Eve can be further processed as follows:

$$\begin{aligned} r_{Ek} &= \mathcal{L}^{-\beta}[\sqrt{P} h_E \mathcal{L}^\alpha[x_k] + n_E)] \\ &= \sqrt{P} h_E \mathcal{L}^{\Delta\alpha}[x_k] + n_E \end{aligned}, \quad (20)$$

$\mathcal{L}^{\Delta\alpha}[x_k]$ is the $\Delta\alpha$-order 4-WFRFT of $x_k$, and the formula is as follows:

$$\mathcal{L}^{\Delta\alpha}[x_k] = \omega_0(\Delta\alpha) \times x_k + \omega_1(\Delta\alpha)\mathcal{L}^1[x_k] + \omega_2(\Delta\alpha)\mathcal{L}^2[x_k] + \omega_3(\Delta\alpha)\mathcal{L}^3[x_k] \quad (21)$$

It can be seen when $\Delta\alpha \neq 0$, Alice transmitted power is divided into four parts, the four parts of the same frequency interference at the same time, only in (21) the first part of $\omega_0(\Delta\alpha) \times x_k$ can be used to correctly demodulate Alice sends information by Eve, the other three parts for Eve, has no effect to correctly demodulate information, It can be regarded as the noise introduced by 4-WFRFT to the eavesdropper.

(20) is expanded as:

$$\begin{aligned} r_{Ek} &= \sqrt{P} h_E \mathcal{L}^{\Delta\alpha}[x_k] + n_E \\ &= \sqrt{P} h_E \{\omega_0(\Delta\alpha) \times x_k + \omega_1(\Delta\alpha)\mathcal{L}^1[x_k] + \omega_2(\Delta\alpha)\mathcal{L}^2[x_k] + \omega_3(\Delta\alpha)\mathcal{L}^3[x_k]\} + n_E \\ &= \sqrt{P} h_E \omega_0(\Delta\alpha) \times x_k + \sqrt{P} h_E \{+\omega_1(\Delta\alpha)\mathcal{L}^1[x_k] + \omega_2(\Delta\alpha)\mathcal{L}^2[x_k] + \omega_3(\Delta\alpha)\mathcal{L}^3[x_k]\} + n_E \end{aligned} \quad (22)$$

Therefore, Eve's instantaneous SNR becomes：

$$\gamma_E = \frac{|\omega_0(\Delta\alpha)|^2 P |h_E|^2}{(1 - |\omega_0(\Delta\alpha)|^2) P |h_E|^2 + N_E}, \quad (23)$$

when Eve the transform order $-\beta$, just meet $\alpha - \beta = 0$, then $|\omega_0(\Delta\alpha)| = 1$, (23) can be reduced to the same as (7), 4-WFRFT has no effect on SNR at eavesdropper.

## III. SECRECY PERFORMANCE ANALYSIS

In this section, assuming that condition of the main channel is known, the joint PDF of the receiving SNR of Bob and Eve after using AS technology in MIMO scenarios is derived, and then the Average security capacity can be derived.

### A. Joint PDF at the receiver

Firstly, receiver SNR joint PDF is derived, under the condition of the Rayleigh fading channel, using the AS of MIMO scenarios, Bob receives PDF of $\gamma_{M,AS}$ can be shown by the following [23]:

$$f_{\gamma_{M,AS}}(z) = N_A N_B [1-\Gamma(1,\frac{z}{\overline{\gamma_M}})]^{N_A N_B -1} \frac{1}{\overline{\gamma_M}} e^{-\frac{z}{\overline{\gamma_M}}}, \quad (24)$$

where $N_A$ is Alice transmitting antenna number, $N_B$ is Bob receiving antenna number, $\overline{\gamma_M}$ is average SNR of legal receiver, $\Gamma(v,z)$ is incomplete Gamma functions, $\Gamma(v,z) \triangleq \int_z^{+\infty} u^{v-1} e^{-u} du$.

By simplifying (24), the PDF of $\gamma_{M,AS}$ is as follows:

$$f_{\gamma_{M,AS}}(z) = \frac{N_A N_B}{\overline{\gamma_M}} e^{-\frac{z}{\overline{\gamma_M}}} \left(1 - e^{-\frac{z}{\overline{\gamma_M}}}\right)^{N_A N_B - 1}, \quad (25)$$

Since Alice and Bob can select the channel with the best channel state through AS, and the transmitting antenna selected by Alice is random to Eve, therefore, the SNR PDF of Eve is the same as the above SISO scenario, which can be expressed in the following form:

$$f_{\gamma_E}(y) = \frac{1}{\overline{\gamma_E}} e^{-\frac{y}{\overline{\gamma_E}}}, \quad (26)$$

According to (25), in MIMO and SISO scenarios, Bob received the relationship between $\gamma_{M,AS}$ and $\gamma_M$ as follows:

$$f_{\gamma_{M,AS}}(z) = N_A N_B F_{\gamma_M}(z)^{N_A N_B -1} f_{\gamma_M}(z) = f_{\gamma_M}(x)|J_1|, \quad (27)$$

where the Jacobian determinant $|J_1|$ can be expressed as follows:

$$|J_1| = \frac{dx}{dz} = N_A N_B F_{\gamma_M}(z)^{N_A N_B -1} \frac{f_{\gamma_M}(z)}{f_{\gamma_M}(x)}, \quad (28)$$

Bob and Eve SNR joint PDF is $f_{\gamma_{M,AS}\gamma_E}(z,y)$ in MIMO scene, Bob and Eve SNR joint PDF is $f_{\gamma_M \gamma_E}(x,y)$ in SISO scene, both satisfy the following:

$$f_{\gamma_{M,AS}\gamma_E}(z,y) = f_{\gamma_M \gamma_E}(x,y)|J_2|, \quad (29)$$

the Jacobian of determinant $|J_2|$ is:

$$|J_2| = \frac{\partial(x,y)}{\partial(z,y)} = \begin{vmatrix} \frac{\partial x}{\partial z} & 0 \\ 0 & 1 \end{vmatrix} = |J_1| = N_A N_B F_{\gamma_M}(z)^{N_A N_B -1} \frac{f_{\gamma_M}(z)}{f_{\gamma_M}(x)}, \quad (30)$$

After obtaining $|J_2|$, the joint PDF of Bob and Eve SNR in the MIMO scene can be obtained:

$$f_{\gamma_{M,AS}\gamma_E}(z,y) = f_{\gamma_M \gamma_E}(x,y) N_A N_B F_{\gamma_M}(z)^{N_A N_B -1} \frac{f_{\gamma_M}(z)}{f_{\gamma_M}(x)} \quad (31)$$

$$= f_{\gamma_M \gamma_E}(z,y) N_A N_B F_{\gamma_M}(z)^{N_A N_B -1}$$

$f_{\gamma_M \gamma_E}(z,y)$ is Bob and Eve received SNR joint PDF in SISO scene, the expression follows as：

$$f_{\gamma_M \gamma_E}(z,y) = \frac{1}{\overline{\gamma_M}\overline{\gamma_E}(1-\rho^2)} e^{-\frac{1}{1-\rho^2}\left(\frac{z}{\overline{\gamma_M}}+\frac{y}{\overline{\gamma_E}}\right)} I_0\left(\frac{2\rho}{1-\rho^2}\sqrt{\frac{z}{\overline{\gamma_M}}\frac{y}{\overline{\gamma_E}}}\right), \quad (32)$$

where $I_0(x)$ is the first class of zero-order modified Bessel function, defined as: $I_0(x) \triangleq \frac{1}{2\pi}\int_0^{2\pi} e^{-x\cos\theta} d\theta$.

Finally, by substituting (32) into (31), the joint PDF of the received SNR of Bob and Eve in the MIMO scenario can be obtained:

$$\begin{aligned}
f_{\gamma_{M,AS}\gamma_E}(z,y) &= f_{\gamma_M \gamma_E}(z,y) N_A N_B F_{\gamma_M}(z)^{N_A N_B -1} \\
&= \frac{N_A N_B}{\overline{\gamma_M}\overline{\gamma_E}(1-\rho^2)} e^{-\frac{1}{1-\rho^2}\left(\frac{z}{\overline{\gamma_M}}+\frac{y}{\overline{\gamma_E}}\right)} I_0\left(\frac{2\rho}{1-\rho^2}\sqrt{\frac{z}{\overline{\gamma_M}}\frac{y}{\overline{\gamma_E}}}\right)\left(1-e^{-\frac{z}{\overline{\gamma_M}}}\right)^{N_A N_B -1} \\
&= \frac{N_A N_B}{\overline{\gamma_M}\overline{\gamma_E}(1-\rho^2)} \sum_{k=0}^{N_A N_B -1} C_{N_A N_B -1}^K (-1)^k e^{-\left[\left(k+\frac{1}{1-\rho^2}\right)\frac{z}{\overline{\gamma_M}}+\frac{1}{1-\rho^2}\frac{y}{\overline{\gamma_E}}\right]} I_0\left(\frac{2\rho}{1-\rho^2}\sqrt{\frac{z}{\overline{\gamma_M}}\frac{y}{\overline{\gamma_E}}}\right)
\end{aligned} \quad (33)$$

where $C_{N_A N_B -1}^K = \frac{(N_A N_B -1)!}{K!(N_A N_B -1-K)!}$.

### B. Average security capacity

In Rayleigh fading channel, the security capacity $C_S$, an important indicator to measure system security, is defined as (34):

$$C_S(\gamma_M, \gamma_E) = [C_M - C_E]^+ = [\log(1+\gamma_M) - \log(1+\gamma_E)]^+, \quad (34)$$

where $[x]^+ = \max\{x,0\}$, which Represent the maximum value between the variable $x$ and 0.

Since the randomness of fading makes Average security capacity $C_S$ a random variable, $C_S$ is usually used for analysis in fading channels. The so-called Average security capacity is the expectation of the instantaneous capacity. The expected value can avoid the fluctuation caused by randomness. In a quasi-static Rayleigh fading channel, the Average security capacity of the system is defined as:

$$\overline{C_S} = E[C_S] = \int_0^\infty \int_0^\infty C_S(x,y) f_{\gamma_M \gamma_E}(x,y) dx dy \quad (35)$$

1) The Average security capacity expression under AS

$$\overline{C_S} = N_A N_B \sum_{k=0}^{N_A N_B -1} C_{N_A N_B -1}^K \frac{(-1)^k}{k+1} \left[\begin{array}{l} e^{\frac{k+1}{\overline{\gamma_M}}} E_1\left(\frac{k+1}{\overline{\gamma_M}}\right) - e^{\frac{1}{\theta_k}\frac{k+1}{\overline{\gamma_E}}} E_1\left(\frac{1}{\theta_k}\frac{k+1}{\overline{\gamma_E}}\right) \\ + \int_0^\infty \frac{e^{-\frac{1}{\theta_k}\frac{k+1}{\overline{\gamma_E}}y}}{1+y} Q\left(A\sqrt{\theta_k y}, B\sqrt{a_k y}\right) dy \\ - \int_0^\infty \frac{e^{-\frac{k+1}{\overline{\gamma_M}}y}}{1+y} Q\left(A\rho\sqrt{y}, B\sqrt{y}\right) dy \end{array}\right],$$

(36)

where $N_A$ is Alice transmitting antenna number, $N_B$ is Bob receiving antenna number; $A \triangleq \sqrt{\frac{2}{1-\rho^2}\frac{1}{\overline{\gamma_M}}}$; $B \triangleq \sqrt{\frac{2}{1-\rho^2}\frac{1}{bc(1-ce^c Ei(-c))}}$; $\theta_k \triangleq k(1-\rho^2)+1$; $a_k \triangleq \frac{\rho^2}{\theta_k} = \frac{\rho^2}{k(1-\rho^2)+1}$; $E_1(x)$ is the exponential integral function, defined as: $E_1(x) = \int_x^\infty \frac{e^{-t}}{t} dt$; $Q(m,n)$ is the first order Marcum Q function, defined as:

$Q(m,n) = \int_n^\infty xe^{-\frac{x^2+m^2}{2}} I_0(mx)dx (m \geq 0, n \geq 0)$.

Average security capacity is defined as the statistical average of instantaneous security capacity. According to (35), $\overline{C_S}$ expression can be written as [24]:

$$\overline{C_S} = \int_0^\infty \int_0^x \frac{1}{1+y} F(y) dy dx, \quad (37)$$

where $F(y)$ an integral variable limit function, which can be obtained by substituting (32):

$$F(y) \triangleq \int_0^y f_{\gamma_M \gamma_E}(x,y)dy$$
$$= \int_0^y \frac{1}{\gamma_M \gamma_E (1-\rho^2)} e^{-\frac{1}{1-\rho^2}\left(\frac{x}{\gamma_M}+\frac{y}{\gamma_E}\right)} I_0\left(\frac{2\rho}{1-\rho^2}\sqrt{\frac{x\,y}{\gamma_M \gamma_E}}\right)dy$$
$$= \frac{1}{\gamma_M} e^{-\frac{x}{\gamma_M}} \int_0^{B\sqrt{x}} v e^{-\frac{v^2+u^2}{2}} I_0(uv)dv \quad (38)$$
$$= \frac{1}{\gamma_M} e^{-\frac{x}{\gamma_M}} \left[1 - Q(A\rho\sqrt{x}, B\sqrt{y})\right]$$

where $u = A\rho\sqrt{x}$, $v = B\sqrt{y}$.

According to (35), in MIMO scenario with AS technology, the expression of security capacity is:

$$\overline{C_S} = \int_0^\infty \int_0^z \frac{1}{1+y} P(y) dy dz, \quad (39)$$

and the relation between $P(y)$ and $F(y)$ is:

$$P(y) = \int_0^y f_{\gamma_{M,AS} \gamma_E}(z,y)dy$$
$$= N_A N_B \sum_{k=0}^{N_A N_B - 1} \binom{N_A N_B - 1}{k}(-1)^k e^{-k\frac{z}{\gamma_M}} \int_0^y f_{\gamma_M \gamma_E}(z,y)dy \quad (40)$$
$$= N_A N_B \sum_{k=0}^{N_A N_B - 1} \binom{N_A N_B - 1}{k}(-1)^k e^{-k\frac{z}{\gamma_M}} F(y)$$

Substituting (40) into (39):

$$\overline{C_S} = \int_0^\infty \int_0^z \frac{1}{1+y} P(y) dy dz$$
$$= N_A N_B \sum_{k=0}^{N_A N_B - 1} \binom{N_A N_B - 1}{k}(-1)^k \int_0^\infty \int_0^z e^{-k\frac{z}{\gamma_M}} \frac{1}{1+y} F(y) dy dz \quad (41)$$
$$= N_A N_B \sum_{k=0}^{N_A N_B - 1} \binom{N_A N_B - 1}{k}(-1)^k \overline{C_{S,k}}$$

next, by substituting (38) into (41), the expression for $\overline{C_{S,k}}$ can be obtained:

$$\overline{C_{S,k}} = \int_0^\infty \int_0^z e^{-k\frac{z}{\gamma_M}} \frac{1}{1+y} F(y) dy dz$$
$$= \int_0^\infty \int_0^z \frac{1}{1+y} \frac{1}{\gamma_M} e^{-(k+1)\frac{z}{\gamma_M}} \left[1 - Q(A\rho\sqrt{z}, B\sqrt{y})\right] dy dz \quad (42)$$
$$= \frac{1}{k+1} e^{\frac{k+1}{\gamma_M}} E_1\left(\frac{k+1}{\gamma_M}\right) - \int_0^\infty \frac{1}{1+y} Q(y) dy$$

where, $Q(y)$ is expressed as:

$$Q(y) = \int_y^\infty \frac{1}{\gamma_M} e^{-(k+1)\frac{z}{\gamma_M}} Q(A\rho\sqrt{z}, B\sqrt{y}) dz \left(\text{make } t = \sqrt{z}\right)$$
$$= \int_{\sqrt{y}}^\infty \frac{2t}{\gamma_M} e^{-(k+1)\frac{t^2}{\gamma_M}} Q(A\rho t, B\sqrt{y}) dt \quad (43)$$

According to (44):

$$\int_\mu^\infty xe^{-\frac{\alpha^2 x^2}{2}} Q(\beta x, \gamma) dx$$
$$= \frac{e^{-\frac{\alpha^2 \mu^2}{2}}}{\alpha^2} Q(\beta\mu,\gamma) + \frac{e^{-\frac{\alpha^2 \gamma^2}{2(\alpha^2+\beta^2)}}}{\alpha^2} - \frac{e^{-\frac{\alpha^2 \gamma^2}{2(\alpha^2+\beta^2)}}}{\alpha^2} Q\left(\mu\sqrt{\alpha^2+\beta^2}, \frac{\beta\gamma}{\sqrt{\alpha^2+\beta^2}}\right) \quad (44)$$

can be obtained:

$$Q(y) = \frac{1}{k+1}\left\{e^{-\frac{k+1}{\gamma_M}y} Q(A\rho\sqrt{y}, B\sqrt{y}) + e^{-\frac{1}{\theta_k}\frac{k+1}{\gamma_E}y}\left[1 - Q(A\sqrt{\theta_k y}, B\sqrt{a_k y})\right]\right\}, \quad (45)$$

By substituting (45) and (42) into (41), it can be obtained that the Average security capacity $\overline{C_S}$ is (36) when AS technology is adopted in MIMO scene.

2) *The Average security capacity expression under 4-WFRFT*

**Theorem 1** In MIMO scenario, using AS and 4-WFRFT technology, the exact expression of Average security capacity is expressed as (46).

$$\overline{C_S} = N_A N_B \sum_{k=0}^{N_A N_B - 1} C_{N_A N_B - 1}^k \frac{(-1)^k}{k+1} \begin{bmatrix} e^{\frac{k+1}{\gamma_M}} E_1\left(\frac{k+1}{\gamma_M}\right) - e^{\frac{1}{\theta_k bc(1-ce^c Ei(-c))}} E_1\left(\frac{1}{\theta_k} \frac{k+1}{bc(1-ce^c Ei(-c))}\right) \\ + \int_0^\infty \frac{e^{-\frac{1}{\theta_k bc(1-ce^c Ei(-c))}y}}{1+y} Q(A\sqrt{\theta_k y}, B\sqrt{a_k y}) dy \\ - \int_0^\infty \frac{e^{-\frac{k+1}{\gamma_M}y}}{1+y} Q(A\rho\sqrt{y}, B\sqrt{y}) dy \end{bmatrix}, \quad (46)$$

where $N_A$ is Alice transmitting antenna number, $N_B$ is Bob receiving antenna number; $A \triangleq \sqrt{\frac{2}{1-\rho^2}\frac{1}{\gamma_M}}$; $B \triangleq \sqrt{\frac{2}{1-\rho^2}\frac{1}{bc(1-ce^c Ei(-c))}}$; $a \triangleq \frac{P\sigma_E^2}{N_E}$; $b \triangleq \frac{|\omega_0(\Delta\alpha)|^2 P\sigma_E^2}{N_E}$; $c \triangleq \frac{1}{a-b} = \frac{N_E}{1-|\omega_0(\Delta\alpha)|^2 P\sigma_E^2}$; $\theta_k \triangleq k(1-\rho^2)+1$; $a_k \triangleq \frac{\rho^2}{\theta_k} = \frac{\rho^2}{k(1-\rho^2)+1}$; $E_1(x)$ is the exponential integral function, defined as: $E_1(x) = \int_x^\infty \frac{e^{-t}}{t} dt$; $Ei(x)$ is the exponential integral function, defined as: $Ei(x) = \int_{-\infty}^x \frac{e^t}{t} dt$; $Q(m,n)$ is the first order Marcum Q function, defined as: $Q(m,n) = \int_n^\infty xe^{-\frac{x^2+m^2}{2}} I_0(mx) dx (m \geq 0, n \geq 0)$.

In (34) of instantaneous security capacity, the instantaneous security capacity is related to the instantaneous SNR, however, in actual communication, we cannot accurately know the instantaneous SNR, so it is impossible to obtain the instantaneous security capacity.in (46), the Average security capacity is ultimately related to the average SNR. In the actual communication process, the average SNR over a period of time can be measured, the Average security capacity can be given by specific parameters over a period of time, therefore, using (46) can effectively evaluate the system security performance.

***Prove*** After using 4-WFRFT, due to 4-WFRFT principle is the constellation of the transmitter and the receiver of modulation and demodulation, the channel conditions haven't changed, according to (6) and (23) two type of analysis can get Bob receiver SNR $\gamma_M$ unchanged and Eve SNR at the receiver $\gamma_E$ is reduced, so as to enhance the safety performance. In the case, the instantaneous security capacity $C_S$ can be expressed as:

$$C_S = \left[\log(1+\gamma_M) - \log(1+\gamma_E)\right]^+$$
$$= \left[\log_2(1+P\frac{|h_M|^2}{N_M}) - \log_2(1+\frac{|\omega_0(\Delta\alpha)|^2 P|h_E|^2}{(1-|\omega_0(\Delta\alpha)|^2)P|h_E|^2 + N_E})\right]^+ \tag{47}$$

In the AS of MIMO scenarios, blending in 4-WFRFT in MIMO scenarios used to supplement the encryption system, the main channel applies 4-WFRFT, legal receiver Bob and transmitter Alice share 4-WFRFT transform order $\alpha$, Eve eavesdropper through technical means to obtain transform order $\alpha$ value, but chose a transform $-\beta$ order as $-\alpha$ estimates, transforming order $-\beta$ for 4-WFRFT.

In MIMO scenario, due to the addition of 4 - WFRFT, decreased the Average security capacity derived from Eve average SNR $\overline{\gamma_E}$ at the receiver of (36) in AS.

According to (23), after 4-WFRFT technology is adopted, the instantaneous SNR of Eve Eavesdropper is:

$$\gamma_E = \frac{|\omega_0(\Delta\alpha)|^2 P|h_E|^2}{(1-|\omega_0(\Delta\alpha)|^2)P|h_E|^2 + N_E}, \tag{48}$$

in (48), $\omega_0(\Delta\alpha)$ is the definite value that changes with parameter $\Delta\alpha$, and $P$ and $N_E$ are constants; In channel modeling, $h_E$ is defined as: $h_E = \sigma_E\left[\left(\sqrt{1-\lambda^2}X_E + \lambda X_0\right) + j\left(\sqrt{1-\lambda^2}Y_E + \lambda Y_0\right)\right]$, where $\sigma_E$ is the fading coefficient of the eavesdropping channel. $X_E, Y_E, X_0$ and $Y_0$ are independent of each other and obey the Gaussian distribution of $\mathcal{N}(0,1/2)$, then $\lambda X_0 \sim \mathcal{N}(0,\lambda^2/2)$, $\sqrt{1-\lambda^2}X_E \sim \mathcal{N}(0,(1-\lambda^2)/2)$ both adding each other, And because the sum of the squares of two independent random variables that obey $\mathcal{N}(0,1/2)$, obey the exponential distribution of exponent 1. Make random variables $T = \left(\sqrt{1-\lambda^2}X_E + \lambda X_0\right)^2 + \left(\sqrt{1-\lambda^2}Y_E + \lambda Y_0\right)^2$, Then the PDF of random variable $T$ is:

$$f_T(t) = \begin{cases} e^{-t} & (t \geq 0) \\ 0 & (t < 0) \end{cases}, \tag{49}$$

the instantaneous SNR of Eve is a function of random variable $T$ as follows:

$$\gamma_E = \frac{|\omega_0(\Delta\alpha)|^2 P|h_E|^2}{(1-|\omega_0(\Delta\alpha)|^2)P|h_E|^2 + N_E} = \frac{bT}{(a-b)T+1}, \tag{50}$$

Application of complex function solution formulas of mathematical expectation of a random variable under the condition of $T(t)$, the average SNR of Eve is equivalent to the mathematical. Therefore, average SNR of Eve at the receiver $\overline{\gamma_E}$ as follows:

$$\overline{\gamma_E} = E[\gamma_E] = \int_0^\infty \gamma_{E(t)}f_{T(t)}dt = \int_0^\infty \frac{bt}{(a-b)t+1}e^{-t}dt$$
$$= \frac{b}{a-b} - \frac{b}{(a-b)^2}e^{\frac{1}{a-b}}\int_0^\infty \frac{e^{-(t+\frac{1}{a-b})}}{t+\frac{1}{a-b}}dt \tag{51}$$

where $a \triangleq \frac{P\sigma_E^2}{N_E}, b \triangleq \frac{|\omega_0(\Delta\alpha)|^2 P\sigma_E^2}{N_E}$.

According to the exponential integral function: $Ei(x) = \int_{-\infty}^x \frac{e^t}{t}dt$, may as well make: $z = t + \frac{1}{a-b} \in (-\infty, -\frac{1}{a-b})$, (51) can be simplified:

$$\overline{\gamma_E} = \frac{b}{a-b} - \frac{b}{(a-b)^2}e^{\frac{1}{a-b}}\int_0^\infty \frac{e^{-(t+\frac{1}{a-b})}}{t+\frac{1}{a-b}}dt$$
$$= \frac{b}{a-b} - \frac{b}{(a-b)^2}e^{\frac{1}{a-b}}Ei(-\frac{1}{a-b}) \tag{52}$$
$$= bc(1-ce^c Ei(-c))$$

where $c \triangleq \frac{1}{a-b} = \frac{N_E}{1-|\omega_0(\Delta\alpha)|^2 P\sigma_E^2}$.

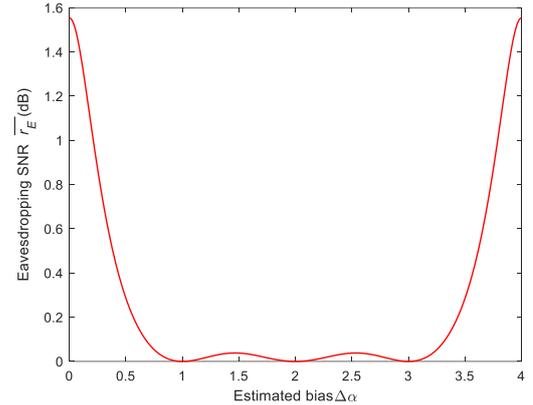

Fig.7. Relationship between Average security capacity and $\Delta\alpha$

By substituting (52) into (36), finally, the expression of Average security capacity $\overline{C_S}$ (46) can be obtained, and ***Theorem 1*** is proved. ∎

Through ***Theorem 1***, after using 4-WFRFT and AS technologies in MIMO scenarios, the average SNR $\overline{\gamma_M}$ of the legal receiver has not been affected by 4-WFRFT technology. However, by comparing (52) and the average SNR of (7), the average SNR $\overline{\gamma_E}$ of the illegal eavesdropper is reduced by 4-WFRFT and depends on the estimated bias $\Delta\alpha$ in Fig.7. In ***Theorem 1***, since the average SNR $\overline{\gamma_M}$ of the legal receiver is unchanged, the average SNR $\overline{\gamma_E}$ of the eavesdropper is reduced, thus, the Average security capacity greatly increased due to the addition of 4-WFRFT.

## IV. SIMULATION ANALYSIS

This section, the Montecarlo simulated $N = 10^6$ times, it is assumed that the main and the eavesdropping channel are subject to static flat Rayleigh fading channel, Rayleigh channel model is set up, by changing the antenna number

$N_A*N_B$, transmitter power $P$, the channel correlation $\rho$, 4-WFRFT transform parameters $\Delta\alpha$, and discussed that AS and 4-WFRFT effect the security performance of the system.

In the following analysis, Table 1 of these initial parameters are constant.

Table 1 INITIAL PARAMETERS

| Parameters | Value |
|---|---|
| Noise power of main channel $N_M$ | -100 dBm |
| Noise power of Eavesdropping channel $N_E$ | -100 dBm |
| Average loss value of Eavesdropping channel $\sigma_E^2$ | -100 dB |
| Average loss value of main channel $\sigma_M^2$ | -95dB |

(1) The relationship between the Average security capacity and the number of antennas

Table 2 INITIAL PARAMETERS OF FIG.8

| Parameters | Value |
|---|---|
| Channel correlation coefficient $\rho$ | 0.5 |
| Transmit power $P$ | 20dBm |
| Estimation error $\Delta\alpha$ | 0 to 4 |

The simulation parameter Settings are shown in Table 2. The simulation verified the influence of the number of antennas $N_A*N_B$ and the change of estimation deviation on the Average security capacity in (46) of theoretical derivation under AS and 4-WFRFT technology, and the results are shown in Fig.8.

It can be seen from Fig.8 that the Montecarlo simulation value is exactly consistent with the exact expression of Average security capacity given by **Theorem 1**, which verifies the correctness of the theoretical derivation of **Theorem 1**. The following conclusions can be draw:

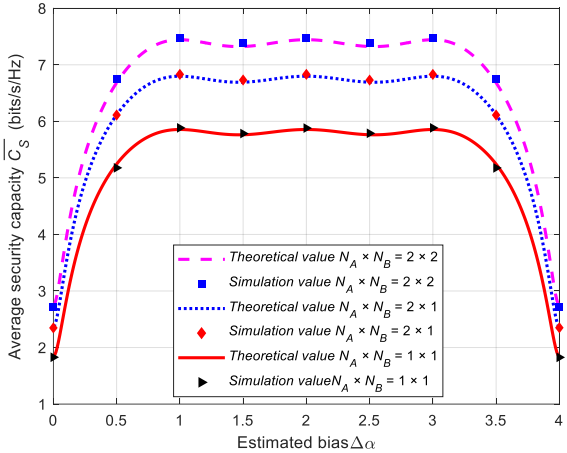

Fig.8. Relationship between Average security capacity and Estimation error

i. When the number of fixed antennas, Fig.8 is about $\Delta\alpha = 2$ symmetrical, so we only discuss $\Delta\alpha \in (0,2)$, $\Delta\alpha \in [2,4]$ in the same way. $\Delta\alpha \in (0,1)$, combining **Theorem 1**, $\Delta\alpha$ of 4-WFRFT have influenced $\omega_0(\alpha)$ weighted parameter in Fig.4, $\omega_0(\alpha)$ goes from 1 to 0, the security capacity of the eavesdropper decreases, while the security capacity of the main channel remains unchanged, therefore, the Average security capacity is significantly improved.

ii. When $\Delta\alpha = 0$, equivalent to Eve accurately estimates Alice's transformation order, 4-WFRFT doesn't improve Average security capacity; When $\Delta\alpha = 1, 2$ and 3, at time $\omega_0(\alpha) = 0$, combining (48), 4-WFRFT decreases illegal Eavesdropper Eve SNR to 0. Therefore, the Average security capacity is maximum at time. When $\Delta\alpha = 1.5$ or 2.5, the Eve the secrecy capacity increases due to $\omega_0(\alpha)$ is improved a little, Average security capacity can reduce a little, but will not reduce too much.

iii. Combined with the exact Average security capacity expression given by **Theorem 1** and Fig.8, the Average security capacity is related to the number of antennas. With the increase of the number of antennas $N_A$ and $N_B$, the Average security capacity will be greatly improved, because the increase of the number of antennas will increase the choice of legitimate communication link channel. In this case, the AS technology can select better communication antenna. But the trend of Average security capacity is the same.

(2) The relationship between the Average security capacity and Channel correlation coefficient

Simulation parameters shows in Table 3 under AS and 4-WFRFT, theoretical derivation of (46), the influence of Channel correlation coefficient and number of antennas $N_A*N_B$ on Average security capacity, the simulation results are shown in Fig.9.

Table 3 INITIAL PARAMETERS OF FIG.9

| Parameters | Value |
|---|---|
| Number of antennas $N_A*N_B$ | 1 and 2 and 4 and 16 |
| Transmit power $P$ | 10dBm |
| Estimation error $\Delta\alpha$ | 1.0 |

i. Through Fig.9, as the channel correlation coefficient $\rho$ is 0 to 1, the average security capacity decreases gradually. Combining **Theorem 1**, because the greater the channel correlation coefficient is, the better the similarity between the main channel and the eavesdropping channel is, which is equivalent to the eavesdropping channel changing its channel condition according to the channel condition of the main channel, so that the channel capacity of the eavesdropping channel is better, leading to the reduction of the average security capacity.

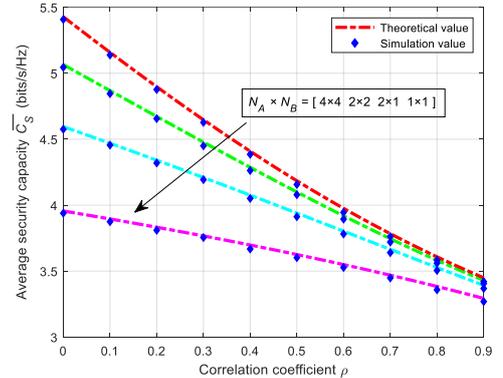

Fig.9. Relationship between Average security capacity and Channel correlation coefficient

ii. Combining **Theorem 1**, Fig. 9 shows that when the channel correlation coefficient is close to 1, the improvement effect of average security capacity is relatively limited. This is because when $\rho$ is large, the similarity between the main channel and the eavesdropping channel is large. When the main channel selects the channel with the best channel state through AS technology, the channel state of the eavesdropping channel will also be significantly improved due to the large channel correlation, resulting in a small increase in the average security capacity.

(3) The relationship between the Average security capacity

and Estimated bias

Simulation parameters shows in Table 4 under AS and 4-WFRFT, theoretical derivation of (46), the channel correlation $\rho$ and the influence of estimation deviation on Average security capacity, the simulation results are shown in Fig.10.

Table 4 INITIAL PARAMETERS OF FIG.10

| Parameters | Value |
|---|---|
| Number of antennas $N_A*N_B$ | 4 |
| Transmit power $P$ | 10dBm |
| Estimation error $\Delta\alpha$ | 0 to 4 |

i. Average security capacity is convergent after $\Delta\alpha=0.9$, so we only analysis $\Delta\alpha\in(0,0.9)$, and symmetrical about $\Delta\alpha=2$, the last analysis of similar. **Theorem 1** shows that the greater the channel correlation $\rho$, the better the main channel and the eavesdropping channel correlation, equivalent to the eavesdropping channel and the main channel of the channel conditions are similar, the secrecy capacity of illegal eavesdropper increased, leading to the Average security capacity decreases.

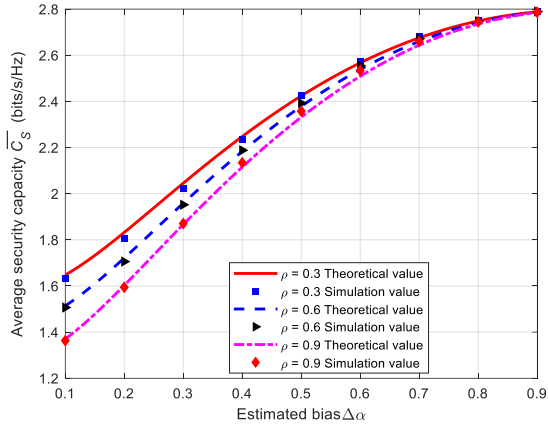

Fig.10. Relationship between Average security capacity and Estimation error and Channel correlation coefficient

ii. When the channel correlation $\rho$ close to 0, the channel between main channel and eavesdropping channel relation is poorer, the actual system, eavesdropper can't obtain the main channel of the channel conditions through technical means, the secrecy capacity of eavesdropper will be reduced, thus improve Average security capacity, to achieve the effect of security.

iii. Fig.10 shows that when $\Delta\alpha$ is 0.1 to 0.9, regardless of the correlation coefficient $\rho\in(0,1)$, Average security capacity converges to a maximum. Because when $\Delta\alpha$ is 0.1 to 0.9, at this time $\omega_0(\alpha)$ is decreased gradually, by **Theorem 1** can shows that Average security capacity related values mainly are influenced by $\omega_0(\alpha)$, eventually when $\omega_0(\alpha)$ is close to 0, eavesdropper get useful information is close to 0, the Average security capacity is the maximum, at this time, the security and confidentiality performance is the best.

(4) The relationship between the Average security capacity and Transmit power

Simulation parameters shows in Table 5 under AS and 4-WFRFT, theoretical derivation of (46), the transmitting power $P$ and the influence of estimation deviation on Average security capacity, the simulation results are shown in Fig.11.

Table 5 INITIAL PARAMETERS OF FIG.11

| Parameters | Value |
|---|---|
| Number of antennas $N_A*N_B$ | 4 |
| Channel correlation coefficient $\rho$ | 0.5 |
| Estimation error $\Delta\alpha$ | 0 to 1 |

i. Fig.11 verifies that the Average security capacity in **Theorem 1** shows a nearly linear growth curve with the increase of transmitting power. According to **Theorem 1**, other parameters are fixed, and the Average security capacity is a linear function of the transmitting power. Therefore, to improve Average security capacity, the transmitting power can be appropriately increased to achieve the purpose of secure communication, but the transmitting power can't be increased without limitation, which will put higher requirements on the transmitter and receiver and lead to more complex system.

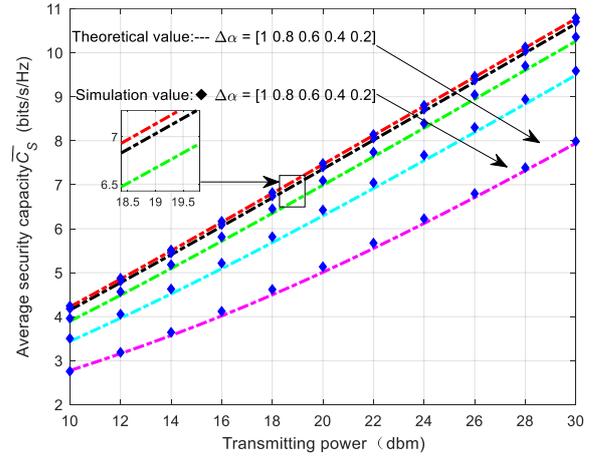

Fig.11. Relationship between Average security capacity and Transmit power

ii. Considering $\Delta\alpha$ is 0 to 1, Average security capacity changes with transmission power. In combination with the results of simulation, we discuss $\Delta\alpha$ is 0 to 1 in **Theorem 1**, equivalent to eavesdropper fails to guess legal the transform order, Eve has useful information that is reduced by 4-WFRFT, thus improving the Average security capacity.

iii. Fig.11 shows that $\Delta\alpha$ is 0 to1, fixed the transmitting power, the difference between Average security capacity decreases gradually. In particular, $\Delta\alpha=0.8$ and $\Delta\alpha=1$, as the transmitting power $P$ increase, Average security capacity is almost overlap, because, Fig.8 shows when $\Delta\alpha=0.8$ to $\Delta\alpha=1$, Average security capacity increase of amplitude is small, so the two curves are almost overlap. But when $\Delta\alpha=0.2$ to $\Delta\alpha=0.5$, the slope is bigger, amplitude is very big, so the difference is bigger than $\Delta\alpha=0.8$ to 1.

## V. CONCLUSION

In this paper, based on the condition of channel correlation, the Rayleigh fading channel model is established in the MIMO scenario, and AS and 4-WFRFT technologies are applied to enhance PLS communication. The exact closed analytic expression of Average security capacity is derived by combining 4-WFRFT and AS innovatively. Through theoretical derivation and the montecarlo simulation analysis get the following conclusion.

Average security capacity $\overline{C_S}$ is related to correlation channel $\rho$, antenna number of $N_A*N_B$, transmit power $P$, and parameter $\Delta\alpha$ in 4-WFRFT. When $\Delta\alpha=0$, Eve accurately estimates Alice transform order, and Average security capacity is minimal and doesn't improve security role. When $\Delta\alpha$ = 1, 2 and 3, Average security capacity is maximum. When $\Delta\alpha \in (0,1)$, due to the security capacity of Eve is reduced, leading to Average security capacity is improved greatly.

In general, 4-WFRFT technology can greatly improve the Average security capacity, and achieve the purpose of increasing the security performance of wireless communication system. Future work will use MIMO algorithm to obtain the Average security capacity without AS technology.